\documentclass[%
reprint,
showpacs
 amsmath,amssymb,
 prd
]{revtex4-1}

\usepackage{graphicx}
\usepackage{hyperref}

\newcommand{\eV}{\textrm{eV}}

\newcommand{\Mpc}{\textrm{Mpc}}
\newcommand{\Gpc}{\textrm{Gpc}}

\newcommand{\s}{\mathrm{s}}
\newcommand{\km}{\mathrm{km}}

\newcommand{\kmax}{k_\mathrm{max}}
\newcommand{\kmin}{k_\mathrm{min}}
\newcommand{\Pgal}{P_\mathrm{gal}}
\newcommand{\PGiggleZ}{P_\mathrm{GiggleZ}}
\newcommand{\Pdamp}{P_\mathrm{damped}}
\newcommand{\Phf}{P_\mathrm{hf}}
\newcommand{\Plin}{P_\mathrm{lin}}
\newcommand{\Pnw}{P_\mathrm{nw}}
\newcommand{\Phfnw}{P_\mathrm{hf,nw}}

\newcommand{\Pcon}{P_\mathrm{con}}
\newcommand{\Pobs}{P_\mathrm{obs}}
\newcommand{\trial}{\mathrm{trial}}
\newcommand{\fid}{\mathrm{fid}}
\newcommand{\zeff}{z_\mathrm{eff}}

\newcommand{\Neff}{N_\mathrm{eff}}
\newcommand{\Hub}{H_0}

\newcommand\eqnref[1]{%
 Eqn.~\ref{eqn:#1}}

\newcommand\figref[1]{%
Fig.~\ref{fig:#1}}

\newcommand\secref[1]{%
Sec.~\ref{sec:#1}}

\newcommand{\f}[1]{{\bf #1}}

\begin{document}

\preprint{APS/123-QED}

\title{The WiggleZ Dark Energy Survey: \\Cosmological neutrino mass constraint from blue high-redshift galaxies}
\author{Signe Riemer--S{\o}rensen$^{1}$, Chris Blake$^{2}$, David Parkinson$^{1}$, Tamara M.\ Davis$^{1}$, Sarah Brough$^3$, Matthew Colless$^3$, Carlos Contreras$^2$, Warrick Couch$^2$, Scott Croom$^4$, Darren Croton$^2$, Michael J.\ Drinkwater$^1$, Karl Forster$^5$, David Gilbank$^6$, Mike Gladders$^7$, Karl Glazebrook$^2$, Ben Jelliffe$^4$, Russell J.\ Jurek$^8$, I-hui Li$^2$, Barry Madore$^{9}$, D.\ Christopher Martin$^5$, Kevin Pimbblet$^{10}$, Gregory B.\ Poole$^2$, Michael Pracy$^{4}$, Rob Sharp$^{3,11}$, Emily Wisnioski$^2$, David Woods$^{12}$, Ted K.\ Wyder$^5$ and H.K.C. Yee$^{13}$}

\affiliation{$^1$School of Mathematics and Physics, University of Queensland, QLD 4072, Australia}\email{signe@physics.uq.edu.au}
\affiliation{$^2$Centre for Astrophysics \& Supercomputing, Swinburne University of Technology, P.O. Box 218, Hawthorn, VIC 3122, Australia}
\affiliation{$^3$Australian Astronomical Observatory, P.O. Box 296, Epping, NSW 1710, Australia}
\affiliation{$^4$Sydney Institute for Astronomy, School of Physics, University of Sydney, NSW 2006, Australia}
\affiliation{$^5$California Institute of Technology, MC 278-17, 1200 East California Boulevard, Pasadena, CA 91125, United States }
\affiliation{$^6$South African Astronomical Observatory, PO Box 9, Observatory, 7935, South Africa}
\affiliation{$^7$Department of Astronomy and Astrophysics, University of Chicago, 5640 South Ellis Avenue, Chicago, IL 60637, United States}
\affiliation{$^8$CSIRO Astronomy \& Space Sciences, Australia Telescope National Facility, Epping, NSW 1710, Australia}
\affiliation{$^{9}$Observatories of the Carnegie Institute of Washington, 813 Santa Barbara St., Pasadena, CA 91101, United States}
\affiliation{$^{10}$School of Physics, Monash University, Clayton, VIC 3800, Australia}
\affiliation{$^{11}$Research School of Astronomy \& Astrophysics, Australian National University, Weston Creek, ACT 2611, Australia}
\affiliation{$^{12}$Department of Physics \& Astronomy, University of British Columbia, 6224 Agricultural Road, Vancouver, BC V6T 1Z1, Canada}
\affiliation{$^{13}$Department of Astronomy and Astrophysics, University of Toronto, 50 St.\ George Street, Toronto, ON M5S 3H4, Canada}

\date{\today}

\begin{abstract}
The absolute neutrino mass scale is currently unknown, but can be constrained by cosmology. The WiggleZ high redshift, star-forming, and blue galaxy sample offers a complementary dataset to previous surveys for performing these measurements, with potentially different systematics from non-linear structure formation, redshift-space distortions, and galaxy bias. We obtain a limit of $\sum m_\nu < 0.60\, \eV$ ($95\%$ confidence) for WiggleZ+Wilkinson Microwave Anisotropy Probe. Combining with priors on the Hubble parameter and the baryon acoustic oscillation scale gives $\sum m_\nu < 0.29\, \eV$, which is the strongest neutrino mass constraint derived from spectroscopic galaxy redshift surveys.
\end{abstract}

\pacs{14.60.Pq, 98.80.Es, 98.62.Py}
\maketitle


\section{Introduction}
Neutrinos are the lightest massive known particles, yet they are treated as exactly massless by the Standard Model of particle physics. {We know they have non-zero masses because neutrino oscillation experiments using solar, atmospheric, and reactor neutrinos have measured mass differences between the three species to be $\Delta m_{32}^2 = 2.43\times10^{-3}\, \eV^2$ and $\Delta m_{21}^2 = 7.59\times10^{-5}\, \eV^2$ \cite{Fukuda:1998,Amsler:2008}. 
The Heidelberg-Moscow experiment has limited the mass of the electron neutrino to be less than $0.35\, \eV$ using $\beta-$spectroscopy \cite{Klapdor:2006}, but no current experiment has sufficient sensitivity to measure the absolute neutrino mass. 

Massive neutrinos affect the way large-scale cosmological structures form by suppressing the gravitational collapse of halos on scales smaller than the free-streaming length at the time the neutrinos become non-relativistic. This leads to a suppression of the small scales in the galaxy power spectrum that we observe today, and consequently we can infer an upper limit on the sum of neutrino masses from the matter distribution of the Universe \cite{Lesgourgues:2006}. Combining with the lower limit of $\sum m_\nu > ~0.05 \eV$ provided by the mass differences from oscillation experiments, allows us to narrow the range of possible neutrino masses.

The cosmic microwave background (CMB) provides an upper limit of $\sum m_\nu < 1.3\, \eV$ \cite[][all limits are 95\% confidence]{Komatsu:2010}. Combining with large-scale structure measurements such as the galaxy power spectrum \cite{Reid:2010,Thomas:2010,Gonzalez-Garcia:2010,dePutter:2012}, galaxy luminosity function \cite{Jose:2011}, cluster mass function \cite{Mantz:2010,Benson:2011}, or the scale of baryon acoustic oscillations \cite[BAO,][]{Reid:2010,Komatsu:2010} tightens the constraints to $\sum m_\nu \lesssim 0.3\, \eV$ by breaking degeneracies with other parameters. Consequently neutrino mass constraints are important goals of current and future galaxy surveys such as e.g.~Baryon Oscillation Spectroscopic Survey \cite{Eisenstein:2011}, Dark Energy Survey \cite{Lahav:2010} and Euclid \cite{Euclid}. In this letter we use the galaxy power spectrum from the WiggleZ Dark Energy Survey to constrain the sum of neutrino masses.

The WiggleZ galaxy survey has several complementary aspects and potential advantages over previous surveys: 
\\1) The neutrino suppression of the galaxy power spectrum is degenerate with effects from non-linear structure formation. Non-linearities increase with time so for the distant galaxies probed by WiggleZ, the contamination from non-linearities is smaller than for previous surveys. This is illustrated in \figref{Pkratio} where we show the ratio between a simulated WiggleZ power spectrum and the linear power spectrum for $z=0.2$ (dashed blue) and $z=0.6$ (solid black). For comparison we also show the ratio for simulated highly-biased massive haloes at $z=0.2$ (dotted red).
\\2) The relationship (bias) between the observed galaxy distribution and the dark matter distribution, which is influenced by massive neutrinos, depends on the observed galaxy type. Previous studies \cite[e.g.][]{Percival:2009,Reid:2010} measured red galaxies, which tend to cluster in the centers of dark matter halos, whereas the star-forming blue WiggleZ galaxies avoid the densest regions. This leads to a lower overall bias, which makes WiggleZ less susceptible to any possible systematics that could arise from a scale-dependence of the bias.

\begin{figure}
	\includegraphics[width=0.99\columnwidth]{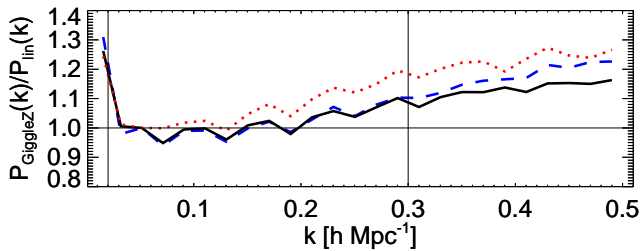}
	\caption{The ratio between the simulated WiggleZ halo power spectrum (from GiggleZ) and the corresponding linear power spectrum (normalised to the first bin) at $z=0.2$ (blue dashed) and $z=0.6$ (black solid) and for luminous red galaxies (dotted red). The vertical lines indicate our fitting range of  $k=0.02-0.3\,h\,\Mpc^{-1}$. Non-linear corrections are clearly less significant for the high-redshift, low-bias WiggleZ halos than at lower redshifts.}
	\label{fig:Pkratio}
\end{figure}

Galaxy redshifts are not entirely determined by the Hubble flow, but also by their peculiar motions (redshift-space distortions) providing a challenge when comparing observations (redshift-space) with theory (real-space). Through exhaustive tests using numerical dark matter simulations of the WiggleZ survey, we demonstrate the breakdown of common models at small scales, and calibrate a new non-linear fitting formula.

\section{The WiggleZ Dark Energy Survey} \label{sec:WiggleZ}
The WiggleZ Dark Energy Survey was designed to detect the BAO scale at higher redshifts than was possible with previous datasets. The $238,000$ galaxies are selected from optical galaxy surveys and ultraviolet imaging by the Galaxy Evolution Explorer to map seven regions of the sky with a total volume of $1\,\Gpc^3$ in the redshift range $z<1$ \cite{Drinkwater:2010}. We split the data into four redshift bins of width $\Delta z=0.2$ with effective redshifts of $\zeff=[0.22, 0.41, 0.6, 0.78]$. The power spectra, $\Pobs$, and covariance matrices, $C$, are measured in $\Delta k = 0.01\,h$\,Mpc$^{-1}$ bins using the optimal-weighting scheme proposed by \citet{Feldman:1994} for a fiducial cosmological model \footnote{$\Omega_b=0.049$, $\Omega_m=0.297$, $h=0.7$, $n_s=1.0$, $\sigma_8=0.8$ \cite{Blake:2010a}}. The Gigaparsec WiggleZ Survey (GiggleZ) simulations \citep{Poole:2012} were designed to probe the low-mass haloes traced by WiggleZ galaxies, whilst providing an equivalent survey volume allowing the measurement of power spectrum modes with $k = 0.01 - 0.5\,h\, \Mpc^{-1}$ and provide a powerful means for testing and calibrating our modeling algorithms. The GiggleZ simulations show that over the range of scales and halo masses relevant for this analysis, the galaxy bias is scale-independent to within 1\% \cite{Poole:2012} whereas the neutrino scale dependent effect is of the order of $5\%$ for a $0.3\, \eV$ neutrino mass \cite{Bird:2011}.


\section{Method} \label{sec:method}
Large-scale structure alone cannot determine all cosmological parameters, so we include data from the CMB as measured by the Wilkinson Microwave Anisotropy Probe (WMAP). To compute the parameter likelihoods, we use importance sampling \cite{Lewis:2002,Swanson:2010} of the WMAP7 Markov Chain Monte Carlo (MCMC) chains available online \footnote{\texttt{http://lambda.gsfc.nasa.gov/product\\/map/dr4/parameters.cfm}}) from fitting to WMAP alone as well as to the chains combining WMAP with the BAO scale from SDSS Luminous Red Galaxies \cite{Percival:2009} and a $H_0 = 74.2\pm3.6\, \km \, \s^{-1} \, \Mpc^{-1}$ prior on the Hubble parameter \cite{Riess:2009} (BAO+$\Hub$). In order to disentangle the neutrino mass and non-linear effects, we test six different approaches for generating a model redshift-space galaxy power spectrum. This section explains the aspects common to all models.

{\bf Matter power spectra:} First we calculate the matter power spectrum, $P_m$, for each redshift bin for the set of cosmological parameters $\mathbf{\theta} = [\Omega_c$ (cold dark matter density), $\Omega_b$ (baryon density), $\Omega_\Lambda$ (dark energy density), $\Omega_\nu$ (neutrino density), $h$ (Hubble parameter), $n_s$ (spectral index), $\Delta^2_R$ (amplitude of primordial density fluctuations)]. We  assume a standard flat $\Lambda$CDM cosmology with no time variation of $w$ in agreement with observational data \cite{Komatsu:2010}. The effective number of neutrinos is fixed to $\Neff=3.04$ assuming no sterile neutrinos or other relativistic degrees of freedom (the 0.04 accounts for the non-thermal nature of the neutrino spectrum, which gets skewed during decoupling because higher-energy neutrinos decouple later than lower-energy neutrinos). The $P_m$ is converted to a galaxy power spectrum, $\Pgal$, using one of the six approaches described in \secref{models}.

{\bf Scaling and convolution:}
Before $\Pgal$ can be compared to the observed power spectrum, $\Pobs$, the survey geometry and the fiducial cosmological model used when measuring the power spectra must be accounted for by Alcock-Paczynski scaling, $a_\mathrm{scl}^3=(D_\mathrm{A}^2 H^{-1}(z))/(D_\mathrm{A, fid}^2 H^{-1}_\fid(z))$, and convolution with the survey window function, $W_{ij}$ \cite{Tegmark:2006,Reid:2009,Swanson:2010}:
\begin{equation}
\Pcon(k_i) = \sum_j \frac{W_{ij}(k)\Pgal(k_j/a_\mathrm{scl})}{a_\mathrm{scl}^3} \, ,
\end{equation}
where $D_A$ is the angular diameter distance and $H(z)$ is the Hubble parameter. Details of the window function can be found in \cite{Blake:2010a}. For all models, we marginalise analytically \cite{Lewis:2002} or numerically over a linear galaxy bias factor.

{\bf Likelihood:}
We assume the power spectra to be distributed as a multivariate Gaussian so the likelihood can be determined as:
\begin{equation}\label{eqn:chi2}
-2\ln (L(\mathbf{\theta})) = \chi^2 = \sum_{i,j}\Delta_i C_{ij}^{-1} \Delta_j \, ,
\end{equation}
where $\mathbf{\theta}$ is the set of cosmological parameters (including galaxy bias), $\Delta_i \equiv [\Pobs(k_i,\mathbf{\theta})-\Pcon(k_i,\mathbf{\theta})]$ with $\Pcon(k_i)$ being the convolved power spectrum in the $i$'th bin, and $C_{ij}$ the covariance matrix. 

The power spectrum measurements in the seven survey regions are treated as independent observations, and their likelihoods combined by multiplication. We require the bias to be the same for all regions at a given redshift. However, we allow the bias to vary between redshifts since the survey magnitude and colour cuts cause the galaxy luminosities to evolve with redshift.

{\bf Importance sampling:}
We use importance sampling to re-weight the WMAP MCMC likelihood chains \cite{Lewis:2002}. For a chain of parameter values $\mathbf{\theta}$ drawn from a likelihood, $L$, it is possible to re-weight the likelihoods with an independent sample from the same underlying parameter distribution. The WMAP and WiggleZ power spectra are independent measurements, so their likelihoods can be combined by multiplication. Thus the weight, $\omega^i_\mathrm{WMAP}$, of each element in the MCMC chain, $i$, can be re-weighted by $\omega^i_{\mathrm{WMAP+WiggleZ}} = L_{\mathrm{WiggleZ}}(\theta^i)\omega_{\mathrm{WMAP}}^i $ \cite{Lewis:2002,Swanson:2010}.

Using the CosmoMC software \footnote{\texttt{http://cosmologist.info/cosmomc/}} for a subsample of the data, we have checked that the preferred regions of parameter space for WMAP and WiggleZ overlap, and consequently importance sampling is a valid method. A CosmoMC module for the WiggleZ power spectra is under development \cite{Parkinson:2012}. We have also fitted the WiggleZ power spectra alone over the range $k=0.02-0.2\,h\,\Mpc^{-1}$ varying only $\Omega_m$ and $f_b$, where $f_b=\Omega_b/\Omega_m$, keeping all other parameters fixed at the WMAP7 best fit values \cite{Komatsu:2010} (WiggleZ alone can not constrain all the cosmological parameters). The resulting parameter values and uncertainties are consistent with the measurements of these parameters using WMAP data alone, which validates the assumption that the two data sets probe the same cosmological parameter space.

{\bf Neutrino mass constraint:}
The neutrino mass limit is calculated from the histogram of the WMAP MCMC chain likelihoods re-weighted by WiggleZ. The X\% confidence upper limit on $\sum m_\nu$ is the value of $M_\nu^{\mathrm{lim}}$ that satisfies:
\begin{equation}
\frac{\sum_{M_\nu^i < M_\nu^{\mathrm{lim}}}L^{i}_{\mathrm{WiggleZ}}(\theta^i)\omega^{i}_{\mathrm{WMAP}}}{\sum_{i=1}^N L^{i}_{\mathrm{WiggleZ}}(\theta^i)\omega^{i}_{\mathrm{WMAP}}} = X/100\, .
\end{equation}
Using the WMAP7 chains alone gives a $95\%$ confidence limit on the neutrino mass of $\sum m_\nu < 1.3\, \eV$ \cite{Larson:2010} and $\sum m_\nu < 0.55\, \eV$ when combining with BAO+$\Hub$ \cite{Komatsu:2010}.

\section{Modeling approaches} \label{sec:models}
Massive neutrinos suppress the power spectrum on all scales smaller than their free-streaming length at the time the neutrinos become non-relativistic. For $\sum m_\nu = 0.3\,\eV$ the most significant suppression happens for $k=0.3-1.3\,h\,\Mpc^{-1}$, but the $k$-dependence of the suppression is most pronounced for $k=0.1-0.3\,h\,\Mpc^{-1}$ and consequently easier to disentangle from other cosmological parameters \cite{Lesgourgues:2006,Bird:2011}. At low redshift structure formation is no longer linear for $k\gtrsim0.1\,h\,\Mpc^{-1}$ \cite{Lesgourgues:2006,Takada:2006,Saito:2008,Scoccimarro:2004}. The standard way of determining the matter power spectrum of non-linear structure formation is the phenomenological Halofit calculation \cite{Smith:2003} distributed with CAMB \footnote{\texttt{http://camb.info}}. Halofit has been derived for massless neutrinos, but using hydro-dynamical SPH simulations \citet{Bird:2011} demonstrated that including realistic neutrinos in the simulations only produced a change in the power spectrum amplitude of less than $1\%$ for $k<0.3\,h\,\Mpc^{-1}$. Simulations show that redshift-space distortions become $k$-dependent at low redshift and consequently are degenerate with neutrino mass \cite{Jennings:2010,Marulli:2011}. 

With the aim of constraining neutrino mass, \citet{Swanson:2010} investigated 12 different models for non-linear structure formation, galaxy bias, and redshift-space distortions. They concluded that models with only one free parameter are unable to provide a good fit for $\kmax \gtrsim 0.1-0.2\,h\,\Mpc ^{-1}$ for the SDSS red and blue galaxies. 
\citet{Blake:2010a} fitted 18 different models to the 2D redshift-space WiggleZ power spectrum for a fiducial cosmology, and concluded that the best fitting models are those of \citet{Jennings:2010} and \citet{Saito:2009} but the latter is very computationally expensive. With these conclusions in mind, we have tested six different approaches for our analysis. The models are fairly similar at low values of $k$, where the large-scale clustering can be treated as linear. There the theory is quite robust, and we expect little difference between the models. However the difference between the models starts to increase for $k>0.2\,h\,\Mpc^{-1}$, which significantly affects the outcome of the fitting. Throughout the analysis we have fixed the lower limit to be $\kmin=0.02\,h\,\Mpc^{-1}$, which corresponds to the largest modes observed in each of the WiggleZ regions (the final results are not very sensitive to the exact value). We then present all results as a function of $\kmax$. The models are shown in \figref{pk} for a fixed cosmology and described below. Testing the models on simulations showed that the models A)-E) are insufficient and the complexity of model F) is necessary.

\begin{figure}
	\includegraphics[width=0.99\columnwidth]{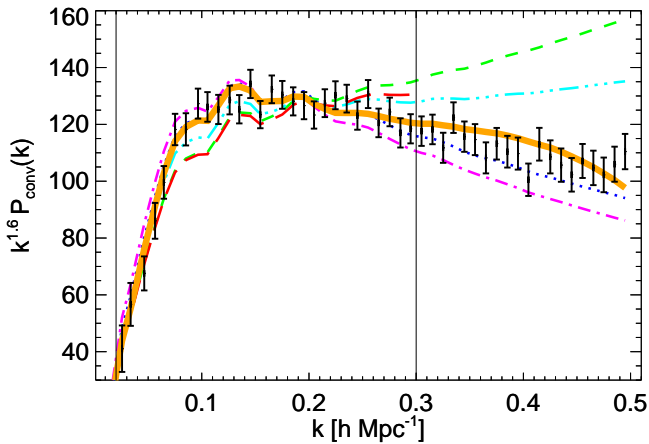}
	\caption{A weighted average of the WiggleZ power spectra in the survey regions and redshifts, and the six models for the best fit cosmology of model F). The models are: A) blue dotted, B) green dashed, C) magenta dot-dashed, D) cyan triple dot-dashed, E) red long dashed, F) thick orange solid. The vertical lines are $k=0.02\,h\,\Mpc^{-1}$ and $k=0.3\,h\,\Mpc^{-1}$. The divergence between the models at large $k$ is clear and demonstrates the importance of careful modeling.}
	\label{fig:pk}
\end{figure}

\f{A) Linear:}
We use CAMB to calculate the linear matter power spectrum, to which we add a linear bias model with redshift-space distortions in the Kaiser limit \cite{Kaiser:1987}.
The model is valid within a few percent for $k\leq 0.15\,h\,\Mpc^{-1}$ \cite{Saito:2009}.

\f{B) Non-linear:}
We use Halofit to calculate the non-linear matter power spectrum, $\Phf$. The bias and redshift-space distortions are treated as for model A).

\f{C) Non-linear with fitting formula for redshift-space distortions and pairwise velocities:}
Combining the ansatz of \citet{Scoccimarro:2004} with fitting formulae derived from simulations, the model of \citet{Jennings:2010} describes the time evolution of the power spectra of dark matter density fluctuations and their velocity divergence field. The details of our implementations of this model are given in Parkinson {\it et al.} \cite{Parkinson:2012}.

\f{D) Non-linear with fitting formula for redshift-space distortions and zero pairwise velocity damping:}
The fitting formulae used in model C) were derived for dark matter particles and not for halos. Setting all galaxy velocity dispersions to zero in model C) provides a better fit to the GiggleZ halo catalogues, and we have treated this special case as a separate model.

\f{E) Non-linear with pairwise galaxy velocity damping:}
Non-linear structure formation leads to increased peculiar galaxy velocities at low redshift, which damps the observed power spectrum. The effect can be described by the empirical model \cite{Peacock:1994}:
\begin{equation}
\Pgal(k) = b_r^2 \Phf(k)\int_0^1\frac{(1+\frac{f}{b_r}\mu^2)^2}{1+(kf\sigma_v\mu)^2}d\mu
\end{equation}
where $f$ is the cosmic growth rate, $\mu = \hat{k}\cdot\hat{z}$ is the cosine of the angle between the wave vector, $\hat{k}$, and the direction of the line of sight, $\hat{z}$, and the one-dimensional velocity dispersion, $\sigma_v$ (in units of $h^{-1}\,\Mpc$), is given by \cite{Scoccimarro:2004}:
\begin{equation} \label{eqn:sigmav}
\sigma_v^2 = \frac{2}{3} \frac{1}{(2\pi)^2} \int d k' \Plin(k')\, .
\end{equation}
Setting $\sigma_v=0\,h^{-1}\,\Mpc$ we recover model B).

\f{F) $N$-body simulation calibrated approach:}
All the non-linear effects are present in an $N$-body simulation for a fiducial cosmology and can be implemented following the approach of \citet{Reid:2010}. For each trial cosmology:
\begin{equation}\label{eqn:GiggleZ}
\Pgal^\trial(k) = b^2\Phfnw^\trial(k)\frac{\Pdamp^\trial(k)}{\Pnw^\trial(k)}\frac{\PGiggleZ^\fid(k)}{\Phfnw^\fid(k)}\, ,
\end{equation}
where 
\begin{equation}
\Pdamp^\trial(k) = \Plin^\trial(k)f_\mathrm{damp}(k)+\Pnw^\trial(k)(1-f_\mathrm{damp}(k))
\end{equation}
and $f_\mathrm{damp}(k)=\exp(-(k\sigma_v)^2)$ with $\sigma_v$ given by \eqnref{sigmav}. $\PGiggleZ^\fid(k)$ is found from a $5^\mathrm{th}$ order polynomial fit to the power spectrum of a set of halos in the GiggleZ simulations chosen to match the clustering amplitude of WiggleZ galaxies. $\Pnw$ and $\Phfnw$ are the power spectra without the acoustic peaks, for the linear and Halofit power spectra respectively. They are calculated from a spline fit to the CAMB power spectra following the approach of \citet{Jennings:2010} and \citet{Swanson:2010}.
The factor of $b^2$ in \eqnref{GiggleZ} is related to galaxy bias. The second factor represents the smooth power spectrum of the trial cosmology. The third factor defines the acoustic peaks and their broadening caused by the bulk-flow motion of galaxies from their initial positions in the density field, and the fourth factor describes all additional non-linear effects in the $N$-body simulation.

\f{Performance of the approaches:}
We tested the different approaches by fitting to the $z=0.6$ power spectrum of a GiggleZ halo catalogue matching the clustering amplitude of WiggleZ galaxies to two sets of 2D parameter grids: $\Omega_m-f_b$, and $\Omega_m-n_s$, with the remaining parameters fixed at the GiggleZ fiducial cosmology values. We chose these grids because the parameters are susceptible to degeneracies with neutrino mass. In both cases we obtain very similar conclusions, so here we only present the results of $\Omega_m-f_b$. For $\kmax<0.2\,h\,\Mpc^{-1}$ most of the models produce a good fit, whereas for $\kmax=0.3\,h\,\Mpc^{-1}$ models B), C) and E) break down and give reduced $\chi^2$ values above $1.5$. The upper panel of \figref{GiggleZchi2} shows the $\chi^2$ for the fiducial GiggleZ cosmological parameters, which is a measure of how well the models recover the input parameters. The lower panel of \figref{GiggleZchi2} shows the difference between $\chi^2$ of the GiggleZ values and the best fit, indicating how far the best fit is from the input values. We assume that the $N$-body simulation, which provides a complete census of the relevant non-linear effects, yields the most accurate clustering model. In this sense the good performance of model F) (\figref{GiggleZchi2}) is a consistency check, and the variations of results produced by the other models are due to the breakdown in their performances compared to the simulation. We are cautious about fitting too small scales where modeling of the non-linearities and massive neutrinos become less robust, and the data is dominated by shot noise. The neutrino implementation in CAMB is only accurate for $\kmax < 0.3\,h\,\Mpc^{-1}$ \cite{Bird:2011} which we take as an upper limit for our fits.

\begin{figure}
	\includegraphics[width=0.99\columnwidth]{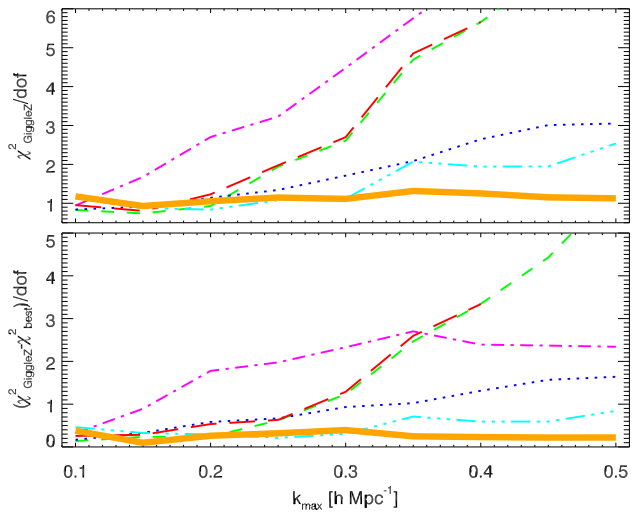}
	\caption{{\it Upper:} Reduced $\chi^2$ of models A)-F) fitted to the $N$-body simulation halo catalogue for the GiggleZ fiducial cosmology values. In absence of systematic errors the models should recover the input cosmology with $\chi^2/\mathrm{dof}=1$. {\it Lower:} Difference in reduced $\chi^2$ values when using the GiggleZ fiducial cosmological parameters and the best fit values. The models are: A) blue dotted, B) green dashed, C) magenta dot-dashed, D) cyan triple dot-dashed, E) red long dashed, F) thick orange solid.}
	\label{fig:GiggleZchi2}
\end{figure}

\section{Results and discussion} \label{sec:results}
When fitting the observed WiggleZ power spectra we obtain the results presented in \figref{kmaxmnuchi2}. The upper panel shows the $\chi^2$ as a function of $\kmax$ for the best fitting parameter values for each of the six approaches, and the lower panel shows the corresponding neutrino mass constraints. 
Although all models produce similar $\chi^2$ values, our comparison with the full $N$-body simulation catalogue (\figref{GiggleZchi2}) revealed that systematic errors arise when models A) to E) are fit across the range of scales $\kmax < 0.3\,h\,\Mpc^{-1}$. Using the fully-calibrated model F), we obtain $\sum m_\nu < 0.60\, \eV$ for WMAP+WiggleZ with $\kmax=0.3\,h\,\Mpc^{-1}$. Combining with BAO+$\Hub$ reduces the uncertainty in $\Omega_m$ and $H_0$, leading to stronger neutrino mass constraints. Without WiggleZ, the WMAP+$\Hub$+BAO dataset gives $\sum m_\nu < 0.55\, \eV$ whereas combining with WiggleZ adds information about the power spectrum tilt ($n_s$). The resulting neutrino mass constraint is  $\sum m_\nu < 0.29\, \eV$ for model F) and $\kmax=0.3\,h\,\Mpc^{-1}$. 

\begin{figure}
	\includegraphics[width=0.99\columnwidth]{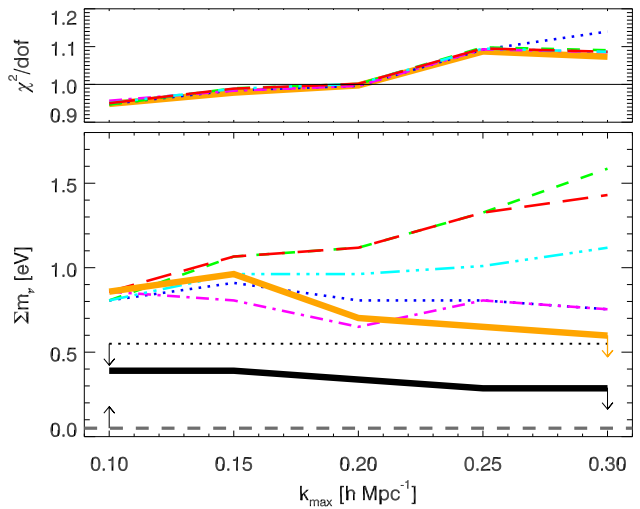}
	\caption{{\it Upper:} Reduced $\chi^2$ as a function of $\kmax$ for each of the six approaches. {\it Lower:} Upper limits on $\sum m_\nu$ as a function of $\kmax$. The models are: A) blue dotted, B) green dashed, C) magenta dot-dashed, D) cyan triple dot-dashed, E) red long dashed, F) thick orange solid.. The dashed grey line is the lower limit from oscillation experiments, and the black lines are upper limits from WMAP+BAO+$\Hub$ (dotted) and WiggleZ+WMAP+BAO+$\Hub$ (solid).}
	\label{fig:kmaxmnuchi2}
\end{figure}

The relative probability distributions of $\sum m_\nu$ for model F) with $\kmax=0.3\,h\,\Mpc^{-1}$ are shown in \figref{prob}. It is clear how adding WiggleZ data to the fit narrows the distributions (dotted to solid) both with (orange) and without (black) the inclusion of BAO+$\Hub$. This is the strongest neutrino mass limit so far derived from spectroscopic redshift galaxy surveys. The advantages of WiggleZ are a higher redshift for which the structure formation is linear to smaller scales, and a simple galaxy bias for the strongly star-forming blue emission line galaxies.

\begin{figure}
	\includegraphics[width=0.99\columnwidth]{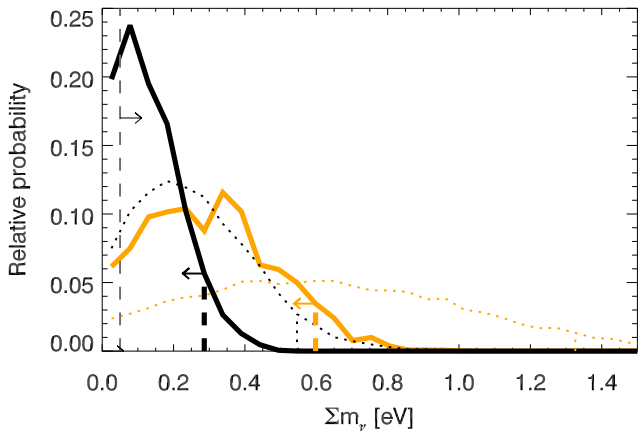}
	\caption{The relative probability distribution of $\sum m_\nu$ from fitting model F) with $\kmax=0.3\,h\,\Mpc^{-1}$ for WMAP (dotted orange), WiggleZ+WMAP (solid orange), WMAP+BAO+$\Hub$ (dotted black) and WiggleZ+WMAP+BAO+$\Hub$ (solid black). The dashed grey line is the lower limit from oscillation experiments, and the vertical lines are 95\% confidence upper limits.}
	\label{fig:prob}
\end{figure}

Our result is comparable to that obtained using photometric redshift galaxy surveys \cite[$\sum m_\nu < 0.28\, \eV$, ][]{Thomas:2010,dePutter:2012}, but the systematics in the two data set are completely different. For example, imaging surveys are potentially susceptible to systematic errors from the imprint of stars on the selection function \cite{Ross:2011} and the shape of the redshift distribution. WiggleZ contains negligible star contamination and much higher redshift resolution compared to photometric surveys. The high redshift and blue galaxies of WiggleZ allow us to fit the power spectrum to smaller scales than previous surveys (both spectroscopic and photometric), where the effect of the neutrinos is larger, and get a similar neutrino mass constraint from a smaller, but well understood galaxy sample. Also the result from galaxy clusters \cite[$\sum m_\nu < 0.33\, \eV$, ][]{Mantz:2010,Benson:2011} is similar, but with different systematics. 

Since the data sets are all independent, they can potentially be combined in the future to provide even stronger constraints. This is particularly interesting in light of recent results  \cite{Komatsu:2010,Kopp:2011} that hint at the existence of additional neutrino species ($N_{\mathrm{eff}}>3.04$). Allowing for additional neutrino species degrades the constraining power of large scale clustering alone, and the combination of $N_\mathrm{eff}$ and $\sum m_\nu$ is therefore poorly constrained with current data.

In the future, galaxy surveys such as the Baryon Acoustic Oscillation Survey, Dark Energy Survey and Euclid will be far more sensitive, giving cosmological neutrino mass constraints of order $\sum m_\nu < 0.05-0.1\, \eV$ \cite{Carbone:2011,Carbone:2011b}. This will be small enough to distinguish between the ordering of the neutrino masses (normal hierarchy where $m_1<m_2<<m_3$ or inverted where $m_3<<m_1<m_2$). However, as demonstrated in this paper, the small details of the modelling of non-linear effects become very important, so robust modelling either theoretically or calibrated to simulations with massive neutrinos will be necessary.

\acknowledgments
{\bf Acknowledgements:} SRS acknowledges financial support from The Danish Council for Independent Research \vline{} Natural Sciences. We acknowledge financial support from the Australian Research Council through Discovery Project grants DP0772084 and DP1093738. SC and DC acknowledge the support of the Australian Research Council through QEII Fellowships.  This research was supported by CAASTRO: http://caastro.org.
GALEX (the Galaxy Evolution Explorer) is a NASA Small Explorer, launched in April 2003. We gratefully acknowledge NASA support for construction, operation and science analysis for the GALEX mission, developed in co-operation with the Centre National d'Etudes Spatiales de France and the Korean Ministry of Science and Technology. We thank the Anglo-Australian Telescope Allocation Committee for supporting the WiggleZ survey over 9 semesters, and we are very grateful for the dedicated work of the staff of the Australian Astronomical Observatory in the development and support of the AAOmega spectrograph, and the running of the AAT.

\bibliographystyle{apsrev4-1.bst}
\bibliography{NeutrinoPRD_draft}


\end{document}